\documentclass[11pt]{article}

\usepackage{a4wide}
\usepackage{natbib}
\usepackage{amsmath, amsthm, amssymb, graphicx}
\usepackage{rotating}
\usepackage{multirow}
\usepackage{url}

\newtheorem{theorem}{Theorem}

\newtheorem{definition}{Definition}

\newtheorem{corollary}[theorem]{Corollary}
\newtheorem{fact}[theorem]{Fact}

\newcommand{\weg}[1]{}
\newcommand{\announce}{\mathcal{L}}
\newcommand{\nl}{\vspace{\baselineskip} \noindent}
\newcommand{\Inter}{\bigcap}
\newcommand{\Union}{\bigcup}
\newcommand{\union}{\cup}

\newcommand{\xset}{\mathcal{Z}}

\title{Avoiding bias in cards cryptography}
\author{M.D.\ Atkinson, H.P.\ van Ditmarsch, and S.\ Roehling}

\begin{document}

\maketitle

\section{Introduction}

Public key cryptography bases its security on mathematical problems that are difficult to solve such as the discrete logarithm problem or factoring the product of two large primes. Advances in technology and new discoveries in mathematics make it more feasible to solve these problems, i.e.\ it becomes more feasible to break the encryption.\footnote{\citet{agrawaletal:2004} show that to determine whether a number is prime can be done in polynomial time. This is not necessarily related to the complexity of determining whether a number is a product of primes.} One solution is to use larger prime numbers to raise the bar a little higher, but this also translates to more computation needed to actually use the encryption, making it more inconvenient for the user. In addition, new mathematical discoveries may suddenly provide an easy way to solve these problems and therefore render the complete algorithm useless for encryption. Instead, one might look at cryptographic protocols in which discovering the secret is not only too complex given the current state of technology but actually impossible. That is, the cryptographic protocol must be developed with a computationally unlimited attacker in mind. One such approach involves the use of a random deck of cards. The general scenario is as follows: Two agents, Alice and Bob, draw $a$ and $b$ cards from a deck of $a+b+c$ cards, and Cathy, the attacker (a.k.a.\ Eve, the eavesdropper), receives the remaining $c$ cards. Alice wishes to communicate her cards to Bob by making a public announcement without informing Cathy of any of her cards. The generalised problem has parameters $(a,b,c)$ and was inspired by what \citet{ditmarsch:russiancards2003} has called the Russian Cards problem, which constitutes the $(3,3,1)$ instance and was presented at the Moscow Mathematical Olympiad in 2000.\footnote{At the time Hans van Ditmarsch did not know that it occurs in \cite{wallis:designs}.} There are also relations to bit-exchange protocols using card deals \citep{fischeretal:1996} and `cards cryptography' as in \cite{stiglic:2001}.

Previous work on the Russian Cards problem involved using epistemic logic to describe its properties and to find its solutions \citep{ditmarsch:russiancards2003,hvd.jancl:2005}, including the analysis of protocols wherein Alice communicates her hand of cards to Bob with more than one announcement, depending on Bob's response to her initial announcement. In \citet{ditmarsch:russiancards2003} it is shown that however Alice structures her announcement, it always corresponds to an announcement of the form ``I hold one of the following hands: \ldots''. The generalised version has also been investigated by \citet*{safecommunication}; they devise combinatorial analogues CA1, CA2 and CA3 of the epistemic requirements for Alice to communicate her hand of cards safely to Bob, and they have found various methods to construct `good announcements' that satisfy these combinatorial axioms.

For the $(3,3,1)$ instance, suppose Alice announces that she holds one of $\{012, 034, 056, 135,$ $246\}$ (by 012 we mean the set $\{0,1,2\}$, etc.). Her announcements are supposed to be truthful, and her actual hand must therefore be among those five. No matter which of those Alice holds, and no matter what Bob holds, he can infer Alice's cards. For example, if Bob held 126, then he could eliminate 012, 056, 135 and 246, leaving only 034, that must therefore be Alice's actual hand. And no matter what Cathy holds, she cannot infer any card of Alice or Bob. Say Cathy held card 5, then she could eliminate 056 and 135, leaving her with 012, 034 and 246 for Alice's hand and (by considering the remaining cards) 346, 126 and 013 for Bob's hand. One can establish that the combinatorial axioms CA1, CA2 and CA3 are indeed satisfied, and that this is therefore a good announcement.

Also for the $(3,3,1)$ instance, and supposing Alice holds 034, she could announce that she holds one of $\{012, 034, 056, 135, 146, 236, 245 \}$. This is a seven-hand announcement. We can similarly establish that this is a good announcement.\footnote{This exact solution is found on the very first page of \cite{wallis:designs}.} 

Some of the constructions proposed by \citeauthor{safecommunication}, while not giving away enough information for Cathy to {\em determine} any card held by Alice or Bob, will result in situations where Cathy can make an {\em educated guess} based on the relative frequency of the cards. For example, consider again the announcement for $(3,3,1)$ given as $\{012, 034, 056, 135, 246\}$. Let us assume that Cathy holds card 3. Then she can exclude all but 012, 056 and 246 from the announcement. Among these, card 2 occurs more often than card 1. In the absence of information on how Alice's announcement has been produced by a protocol, and discounting that Alice (of course) will anticipate such a line of argument, it is reasonable for Cathy to assume that all of Alice's holdings are equally likely. But in that case Alice is more likely to hold card 2 than card 1. In other words: even though the announcement is unbiased with respect to holdings, the announcement may be biased with respect to card occurrences, or otherwise biased with respect to patterns in the announcement, and this information may be valued by the eavesdropper (attacker) Cathy. The other announcement, $\{012, 034, 056, 135, 146, 236, 245 \}$, does not contain card occurrence bias.

From the perspective of cryptography, there appear to be two distinct ways to overcome such bias: {\em either} use protocols that produce announcements that are unbiased for card occurrence (or more complex patterns), {\em or} use protocols that (may) produce biased announcements but ensure that there is no relation between patterns in the announcement, such as card occurrence, and the actual holding. We focus on the first, in Section \ref{uannounce}, using design theory. We devise an additional requirement for the announcement in order to eliminate the possibility of making educated guesses. To that effect we propose an additional combinatorial axiom CA4. We give a method to design announcements that meet this requirement, {\em unbiased announcements} therefore, and we prove some relevant results. Those are our main contributions. Additionally, in Section \ref{uprotocol}, we present {\em unbiased protocols} for the $(3,3,1)$ case, counteracting single card occurrence bias in announcements.

\section{Unbiased announcements} \label{uannounce}

\subsection{Combinatorial axiom CA4}

We will use terminology as in \citet{safecommunication}. Cards are commonly referred to as points, are all distinct, and are labeled with consecutive natural numbers. The set of all cards (or deck of cards) is denoted by $\Omega$. An {\em $i$-set} is a set of $i$ cards. A possible holding (or hand) of Alice is called a {\em line} (in other words, a line is an $a$-set). Thus, an announcement $\announce$ by Alice consists of one or more lines. We write $X,Y,Z$ for $i$-sets, $x,y,z$ for points in such sets, and in particular also $L$ for $a$-sets (lines). Alice, Bob, and Cathy hold, respectively, $a,b$, and $c$ cards. These are the parameters of the card deal, for which we write $(a,b,c)$. `Elimination' refers to Cathy or Bob eliminating those lines from the announcement that are impossible holdings for Alice because they contain one or more of their own cards.

\citeauthor{safecommunication} proposed three axioms CA1, CA2, and CA3, that correspond to the informal requirements given in the problem description for Alice to inform Bob of her cards. An announcement satisfying those axioms is called a {\em good announcement}. A good announcement guarantees that it is common knowledge among Alice, Bob, and Cathy that Bob knows Alice's holding. The axioms are as follows (CA stands for `Combinatorial Axiom'). 
\begin{itemize}
\item[CA1] For every $b$-set $X$ there is at most one line in $\announce$ that avoids $X$.
\item[CA2] For every $c$-set $X$ the lines in $\announce$ avoiding $X$ have empty intersection.
\item[CA3] For every $c$-set $X$ the lines in $\announce$ avoiding $X$ have union consisting of all cards except those of $X$. 
\end{itemize}

Combinatorial Axiom 1 states that, given the announcement, Bob must be able to infer what Alice is holding. In order for Bob to figure out which line of the announcement is Alice's holding, he has to eliminate lines from the announcement based on his knowledge of his own cards. For example, because cards are distinct, if Bob holds card 4, then he can eliminate all lines that contain card 4 since those cannot be a possible holding of Alice. Similarly, Bob can eliminate any other line that contains a card that he himself holds. A line in the announcement that contains none of the cards held by Bob is said to avoid Bob's hand (here denoted by $b$-set $X$). If there are two or more such lines in the announcement, then Bob is left with more than one possibility for Alice's hand and cannot state with certainty which is the correct one. Therefore, there should be at most one line in the announcement that avoids Bob's hand. (As we are assuming that the announcement is truthful and that Alice's hand is among the lines, there is even {\em exactly} one line that avoids Bob's hand.)

Combinatorial Axiom 2 states that, given the announcement, Cathy must not be able to infer {\em any} card held by Alice.  Cathy employs the same process of eliminating lines from the announcement as Bob by looking at her own hand (denoted by $c$-set $X$). After elimination, she examines the remaining lines. If there is one card common to all these lines, then Cathy can conclude that Alice holds that card. So, there must be no card common to all remaining lines. In other words, all remaining lines taken together must have empty intersection.  

Combinatorial Axiom 3 states that, given the announcement, Cathy must not be able to infer {\em any} card held by Bob. If it is not satisfied, there is a card that does not occur among the lines avoiding Cathy's holding $X$. This card is therefore not held by Alice, nor is it held by Cathy. It must therefore be a card held by Bob.

For parameters $(3,3,1)$, the announcements $\{012, 034, 056, 135, 246 \}$ and $\{012, 034, 056,$ $135,$ $146, 236, 245 \}$ both satisfy CA1, CA2, and CA3, as can be easily checked. We propose to distinguish between these announcements by means of another, new, combinatorial axiom. This is CA4. It expresses absence of card occurrence bias. We also propose yet another axiom, CA5, that will then be shown equivalent to CA4.

\begin{itemize}
\item[CA4] For every $c$-set $X$ there is a number $n_X$ such that for every point $x \notin X$ there are $n_X$ lines in $\announce$ avoiding $X$ that contain $x$. 
\item[CA5] For every $c$-set $X$ there is a number $m_X$ such that for every point $y \notin X$ there are $m_X$ $b$-sets $Y$ avoiding $X$ that contain $y$ and that avoid an $L \in \announce$ also avoiding $X$. 
\end{itemize}
Combinatorial Axiom 4 states that, given Alice's announcement and Cathy's hand of cards, no card occurs more often than another one in the lines Cathy considers possible. Combinatorial Axiom 5 states that, given Alice's announcement and Cathy's hand of cards, no card occurs more often than another one in the $b$-sets Cathy considers possible for Bob. 

The new combinatorial axioms become more readable if we introduce additional formalisation. Given a collection $\xset$ of $i$-sets $Z \subseteq \Omega$ (lines, $b$-sets, $c$-sets, ...), the subset of $\xset$ with all points contained in $X \subseteq \Omega$ is denoted $\xset(X)$, i.e.\ \[ \xset(X) = \{ Z \in \xset \mid Z \subseteq X \}. \] On the other hand, the set of $i$-sets in $\xset$ containing (all) points in $X$ is denoted $\xset[X]$, i.e.\ \[ \xset[X] = \{ Z \in \xset \mid X \subseteq Z \}. \] For $\xset(\{x\})$, write $\xset(x)$, and for $\xset[\{x\}]$, write $\xset[x]$; for $\xset(X \union \{x\})$ we write $\xset(X + x)$, for $\xset( \{x,y\})$ we write $\xset(xy)$, etc. The complement of $X$ in $\Omega$ is $\overline{X}$. We combine the notations, e.g.\ we write $\announce(\overline{X})[x]$ for the set of lines in $\announce$ avoiding $X$ and containing $x$. Finally, somewhat arbitrarily, $b(\announce(\overline{X}))$ is the set of $b$-sets $Y$ avoiding $X$ and an $L \in \announce$ also avoiding $X$, i.e.\ \[ b(\announce(\overline{X})) = \{ Y \mid Y = \Omega - X - L, L \in \announce, L \cap X = \emptyset \}. \] We now can rephrase the combinatorial axioms as
\begin{definition}[Combinatorial Axioms] We distinguish five axioms.
\begin{itemize}
\item[{\bf CA1}] For every $b$-set $X$: $|\announce(\overline{X})| \leq 1$.
\item[{\bf CA2}] For every $c$-set $X$: $\Inter \announce(\overline{X}) = \emptyset$.
\item[{\bf CA3}] For every $c$-set $X$: $\Union \announce(\overline{X}) = \overline{X}$.
\item[{\bf CA4}] For every $c$-set $X$ there is a number $n_X$ such that for every $x \notin X$: $|\announce(\overline{X})[x]| = n_X$.
\item[{\bf CA5}] For every $c$-set $X$ there is a number $m_X$ such that for every $x \notin X$: $|b(\announce(\overline{X}))[x]| = m_X$.
\end{itemize}
\end{definition}

Announcement $\{012, 034, 056, 135, 246 \}$ does not satisfy CA4. Take $X = \{5\}$. The lines not containing 5 (i.e., avoiding $\{5\}$) are $012$, $034$ and $246$. {\em Two} of those contain 2 but only {\em one} line contains 1. Therefore, no number $n_5$ (i.e., $n_{\{5\}}$) exists in this case. On the other hand, announcement $\{012, 034, 056, 135, 146, 236, 245 \}$ satisfies CA4, with $n_y = 2$ for all points $y = 0,\dots,6$. E.g., $\{135, 146, 236, 245 \}$ avoid 0; point 1 occurs twice in those, namely in 135 and 146; and so on for other points. Announcement $\{012, 034, 056, 135, 246 \}$ does not satisfy CA5. Take $X = \{5\}$. The $b$-sets not containing 5 and avoiding one of $012, 034$ and $246$ are: $346, 126$ and $013$. Two of those contain a 1 but only one contains a 2. Again, the seven-line announcement satisfies CA5. 

Many other, and more generic, examples can be found using design theory (see \citet{wallis:designs,hughes:tdesigns}). The mathematical theory of block designs deals with collections of special subsets, called blocks (or lines), of a given set. It provides a convenient framework for studying the relation between the proposed combinatorial axioms CA4 and CA5. A $t$-design with parameters $(v,k,\lambda)$ has the property that any combination of $t$ distinct elements of a set of $v =|\Omega|$ points occurs in the same number $\lambda$ of $k$-blocks (or $k$-lines). The number $\lambda$ is referred to as the covalency of the design. Thus, in $2$-designs, also known as balanced incomplete block designs, any pair of distinct cards occurs in the same number of lines. This is relevant for our investigation, because it entails that in the subset of lines containing any given card (such as a singleton $c$-set), any other card occurs in the same number of lines. Similarly, in $3$-designs any $3$-tuple of distinct cards occurs in the same number of lines. This can be further generalised to $4$-designs, $5$-designs, etc., but such designs are far less common and few general constructions are available that may help us here. Every $t$-design is also a $1$-design, $2$-design, \ldots, $(t-1)$-design. The seven-hand announcement $\{012, 034, 056, 135, 146, 236, 245 \}$ is a 2-design, with block size 3 and covalency 1. CA4 can be formulated as \begin{quote} For every $c$-set $X$, $\announce(\overline{X})$ is a 1-design with covalency $n_X$. \end{quote}

We can construct designs satisfying CA4 using the various methods known for constructing designs, such as from projective planes and binary designs. Incidental results are reported in \citet{yates} and \citet{bose}. For details, we refer to \citet{roehling}. Here, we only show how binary designs can be used to construct announcements satisfying CA4. 

\paragraph*{Binary designs} Binary designs give solutions for $a=2^{n-1}, b=a-1$, and $c=1$, for $n \geq 3$. Here, $n$ is the number of bits used in the construction. These designs are special because the same $n$ may be associated with more than one instance of the $(a,b,c)$ parameters. For example, $(8,7,1)$ (satisfying the above for $n=4$) and $(8,6,2)$ have the same solution given by a binary design with $n=4$. In Theorem \ref{thm:binary}, later, we prove that binary designs are $3$-designs.\footnote{Specifically, they are 3-$(2^n, 2^{n-1},1)$ designs.} This is sufficient to guarantee that CA4 is satisfied when $c=1$, using another result, Theorem \ref{thm:CA4implies2}.

Binary designs are constructed as follows. Choose a number of bits $n \geq 3$. For all $2^n-1$ $n$-bit vectors $(y_1, y_2, ..., y_n)$ (except all zeros) solve the equation 
$x_1y_1 + x_2y_2 + ... + x_ny_n = 0$, 
where $x_i=0,1$. There are $2^{n-1}$ solutions to each equation, each $x_1x_2...x_n$ representing a point in binary. The points gained from an equation together constitute a line. This produces $2^n-1$ lines (one per equation). For each line, compute the complement by taking all binary points that are {\em not} present in the line; these complements are also taken as lines. Now we have a total of $2(2^n-1)$ lines which constitute the announcement. To get the final announcement using our format, replace every point in binary with its decimal representation. 

For an example, we construct a binary design with $n=3$.\footnote{These are originally known as Steiner quadruples \cite[p.71]{colbournetal:1996}.} Each line consists of $2^{n-1}=4$ points. The $2^n-1=7$ non-zero $3$-bit vectors are $001, 010, 011, 100, 101, 110, 111.$ The two lines corresponding to the first vector are $\{000, 010, 100, 110\}$ (in decimal notation $\{0,2,4,6\}$, i.e.\ 0246) and $\{001, 011, 101, 111\}$ (in decimal 1357). Proceed similarly for the remaining $3$-bit vectors. The resulting announcement consisting of the $2(2^n-1)=14$ lines is
\[ \{ 0246, 0145, 0347, 0123, 0257, 0167, 0356, 1357, 2367, 1256, 4567, 1346, 2345, 1247 \} \]
Given parameters $(4,3,1)$, this announcement $\announce$ satisfies CA4: for all points $y$, $n_y = 4$. For example, for $y=0$ we get $\announce(\overline{0}) = \{4567, 2367, 2345, 1357, 1346, 1256, 1247\}$ and all other points occur exactly four times in this set: point 1 in the last four lines, point 2 in lines 2, 3, 6, and 7; etc.

\nl Apart from CA4, which for parameters $(a,b,1)$ amounts to checking whether $\announce(\overline{x})$ is a $1$-design for arbitrary $x$, one could imagine strengthening the requirements, for example, demand that $\announce(\overline{x})$ is a 2-design for all points $x$ as well. We have already seen that the seven-hand announcement for $(3,3,1)$ also satisfies this requirement. We will feature a result for this stronger requirement in the next section, in Theorem \ref{himike}.

\subsection{Theoretical results}

\begin{theorem}\label{thm:ca4ca5}
CA4 if and only if CA5.
\end{theorem}
\begin{proof} 
Assume CA4 holds. Let $X$ be any set of $c$ points. For every line $L$ in $\announce(\overline{X})$ there is a $b$-set $\Omega - X - L$ in $b(\announce(\overline{X}))$. Therefore, $|b(\announce(\overline{X}))| = |\announce(\overline{X})|$. Also, for all points $y\in X$, if $y \in Y \in \announce(\overline{X})$ then $y \not \in Z \in b(\announce(\overline{X}))$ where $Z = \Omega - Y - X$. Point $y$ occurs in $n_X$ lines in $\announce(\overline{X})$. It therefore does {\em not} occur in $n_X$ lines in $b(\announce(\overline{X}))$, and it therefore occurs in $|b(\announce(\overline{X}))| - n_X$ lines in $b(\announce(\overline{X}))$. As this is for arbitrary $y$, this defines the number $m_X$. The argument runs both ways.
\end{proof} 
In other words, we can forget about CA5 from here on.
\begin{theorem}\label{thm:CA4implies2}
Let $c=1$. (CA4 holds and $\announce$ is a $1$-design) if and only if $\announce$ is a $2$-design. 
\end{theorem}
\begin{proof}
Assume $\announce$ is a $1$-design and that CA4 holds, and that $c=1$. CA4 says that every $\announce(\overline{x})$ is a $1$-design. Its size $|\announce(\overline{x})|$ is independent of $x$, as $\announce$ is a $1$-design. For arbitrary $y \in \announce(\overline{x})$, $|\announce(\overline{x})[y]| + |\announce[x+y]| = |\announce[y]|$ (note that $|\announce[x+y]| = |\announce[x][y]|)$. As $|\announce(\overline{x})[y]|$ and $|\announce(y)|$ are independent of $x$ and $y$ (by CA4, and because $\announce$ is a 1-design, respectively), so is $|\announce[x + y]|$. Therefore $\announce$ is a 2-design.

Assume $\announce$ is a $2$-design, i.e. $|\{L \in \announce \mid x,y \in L \}| = \lambda_2$ is independent of $x$ and $y$. We want to show that $|\{L \in \announce(\overline{x}) \mid z \in L\}| = n_x$ is independent of $z$, for any holding $x$ of Cathy. Note that when she eliminates lines from $\announce$ that contain $x$, she reduces the number of lines containing any $y \neq x$ by $\lambda_2$. Let $\lambda_1 = |\{ L \in \announce \mid y \in L \}|$, which is independent of $y$ because $\announce$ is also a $1$-design. Before elimination $\lambda_1$ lines contained $y$. After elimination, $\lambda_1 - \lambda_2$ lines contain $y$. This is the number of lines $n_x$ in $\announce(\overline{x})$ that contain $y$. Since it is independent of $y$, CA4 holds. 
\end{proof}

\begin{theorem}\label{thm:nx}
If CA4 holds then $n_X = \frac{a|\announce(\overline{X})|}{a+b}$.
\end{theorem}
\begin{proof}
Count the total number of cards occurring in $\announce(\overline{X})$ in two ways. Assuming CA4 holds there are $a+b$ distinct cards and each of them occurs $n_X$ times. There are $|\announce(\overline{X})|$ lines and each of them contains $a$ cards. Thus $(a+b)n_X = a|\announce(\overline{X})|$. 
\end{proof}

\begin{theorem}\label{thm:nxindep}
Let $c=1$. If CA4 holds then $n_X$ is independent of $X$.
\end{theorem}
\begin{proof}
Assume $c=1$ and CA4 holds. Take two arbitrary distinct $X_1 = \{x_1\}$ and $X_2 = \{x_2\}$. Consider $\announce(\overline{x_1})$. It contains no lines that contain card $x_1$ and $n_{x_1}$ lines that contain card $x_2$. It must therefore contain $|\announce(\overline{x_1})| - n_{x_1}$ lines that contain neither card $x_1$ nor card $x_2$. And due to construction of the set, this is the exact number of lines in $\announce$ that contain neither card. Now consider $\announce(\overline{x_2})$. It contains no lines that contain card $x_2$ and $n_{x_2}$ lines that contain card $x_1$. It must therefore contain $|\announce(\overline{x_2})| - n_{x_2}$ lines that contain neither card $x_1$ nor card $x_2$. And due to construction of the set, this is the exact number of lines in $\announce$ that contain neither card. Thus, we get the following equation 
\begin{eqnarray*}
|\announce(\overline{x_1})| - n_{x_1} &=& |\announce(\overline{x_2})| - n_{x_2} \\
n_{x_1}\frac{a+b}{a} - n_{x_1} &=& n_{x_2}\frac{a+b}{a} - n_{x_2} \\
n_{x_1}\left(\frac{a+b}{a}-1\right) &=& n_{x_2}\left(\frac{a+b}{a}-1\right) \\
n_{x_1}\left(\frac{b}{a}\right) &=& n_{x_2}\left(\frac{b}{a}\right) \\
n_{x_1} &=& n_{x_2}
\end{eqnarray*}
Because $x_1$ and $x_2$ were chosen arbitrarily we conclude that $n_x$ is independent of $x$.
\end{proof}
The following follow directly from Theorems \ref{thm:nx} and \ref{thm:nxindep}.
\begin{corollary}
If CA4 holds, then $|\announce(\overline{x})|$ is independent of $x$.
\end{corollary}
\begin{corollary}
If CA4 holds, $|\announce(\overline{x})|$ is independent of $x$ if and only if $n_x$ is independent of $x$.
\end{corollary}

\begin{theorem}\label{thm:binary}
Binary designs are $3$-designs.\footnote{This is an original proof. But we find it likely that this is a known result in design theory. We did not find a reference to binary designs as an infinite family of $3$-designs in either \cite{wallis:designs} or \cite{colbournetal:1996}, except for the already mentioned specific case for $n=3$: Steiner quadruples.} 
\end{theorem}
\begin{proof}
By construction binary designs are based on $2^n$ points and $2(2^n-1)$ lines of size $2^{n-1}$. 
We now show that any three points $p_1, p_2, p_3$ occur in exactly $2^{n-2} -1$ lines.
The lines of the design contain points of the form $x_1x_2 \ldots x_n$, that is, a point is represented by an $n$-vector ${\bf x}$. For any $n$-vector ${\bf y} \neq {\bf 0}$ there is a line $ \{ {\bf x} \mid {\bf y}\cdot{\bf x} = 0 \}$ which we call $bx_0$ and a line $\{ {\bf x} \mid {\bf y}\cdot{\bf x} = 1 \}$ which we call $bx_1$.
Let the $n$-vectors ${\bf u}, {\bf v}, {\bf w}$ be three distinct points in these lines. A line containing these comes from a vector ${\bf y}$ with either ${\bf y} \cdot {\bf u} = 0$ and ${\bf y} \cdot {\bf v} = 0$ and ${\bf y} \cdot {\bf w} = 0$, in the case of $bx_0$ (equation $i$); or ${\bf y} \cdot {\bf u} = 1$ and ${\bf y} \cdot {\bf v} = 1$ and ${\bf y} \cdot {\bf w} = 1$, in the case of $bx_1$ (equation $ii$). We now have two cases:
\begin{enumerate}
	\item One of ${\bf u}, {\bf v}, {\bf w}$ is the sum of the other two. Then ${\bf u}, {\bf v}, {\bf w}$ form a subspace $U$ of the vector space $V$ of dimension $n$ over the field of $2$ elements. The subspace $U$ has dimension 2. Note that the $bx_0$'s come from $U^\perp = \{ {\bf y} \mid {\bf y} \cdot {\bf t} = 0 \textrm{ for all } {\bf t} \in U \}$ and that $\dim{U} + \dim{U^\perp} = n$. Therefore, the dimension of $U^\perp$ is $n-2$ and the number of $bx_0$'s is $2^{n-2}-1$ (excluding ${\bf y}={\bf 0}$). The number of $bx_1$'s is 0 because Equation $ii$ cannot hold. 
	\item The vectors ${\bf u}, {\bf v}, {\bf w}$ are linearly independent. 
	Then ${\bf u}, {\bf v}, {\bf w}$ form a subspace of dimension 3 and by the same argument as above the number of $bx_0$'s is $2^{n-3} - 1$. The number of $bx_1$'s is $2^{n-3}$ because we just find one ${\bf b_0}$ for which Equation $ii$ holds and then add all the $bx_0$'s. 
\end{enumerate}
In both cases the number of lines that contain the three points is $2^{n-2}-1$ in total. Thus, binary designs are $3$-designs. 
\end{proof}

\begin{corollary}
A binary design will satisfy CA4 for $c=1$.
\end{corollary}
\begin{proof}
Directly from Theorem \ref{thm:CA4implies2} and Theorem \ref{thm:binary}.
\end{proof}

\begin{theorem} \label{himike}
$\announce$ is a 3-design if and only if $\announce(\overline{x})$ is a 2-design for all points $x$ and $\announce$ is a $1$-design.
\end{theorem} 
\begin{proof}
Let $\announce$ be a $3$-design. Trivially, it is also a $2$-design, and also a $1$-design. The last satisfies one proof obligation. As $\announce$ is a $3$-design, $|\announce[yzx]|$ is independent of $y,z,x$. As $\announce$ is a $2$-design, $|\announce[yz]|$ is independent of $y,z$; and therefore independent of $y,z,x$ (note that $x$ does not occur at all in $|\announce[yz]|$). As $|\announce[yz]|=|\announce[yzx]|+|\announce[yz](\overline{x})|$, also $|\announce[yz](\overline{x})|$ is independent of $y,z,x$, i.e., for every $x$, $\announce(\overline{x})$ is a $2$-design.

Assume $\announce(\overline{x})$ is a $2$-design for all $x$, and that $\announce$ is a $1$-design. Similarly to above it immediately follows that $\announce$ is a $3$-design. 
\end{proof}

\begin{corollary}
Binary designs $\announce$ satisfy that $\announce(\overline{x})$ is a 2-design for all points $x$.
\end{corollary}
\begin{proof}
Directly, from Theorems \ref{thm:binary} and \ref{himike}.
\end{proof}

\section{Unbiased protocols} \label{uprotocol}

In the previous section we focussed on avoiding bias in an announcement. Such bias resulted from the overrepresentation of certain patterns, such as single cards, or pairs of cards, or triples, in the announcement or in the remaining lines avoiding a given $c$-set (eavesdropper Cathy's hand of cards). Announcements where arbitrary $c$-set avoiding lines always are 1-designs, or 2-designs, or 3-designs (respectively), guarantee that such bias is absent. The suggested link between overrepresentation of patterns (such as individual card occurrence) in an announcement and the probability of that pattern occurring in the actual holding is, of course, that {\em each line in an announcement is equally likely to be the actual holding}. Given an underlying protocol to produce such an announcement, this is achieved when each announcement resulting from the protocol's execution is equally likely to be produced. In the absence of information to the contrary, that may be a reasonable assumption.

But another way to avoid bias in cryptographic communication is to apply a protocol that takes such overrepresentation of patterns in announcements into account. By making that protocol public, the sender and receiver can then unbias the announcement---but just as well they may keep it secret, and in that case have a cutting edge over an unsuspecting eavesdropper. In other words, by applying protocols that make some lines in an announcement more likely to be the actual holding than others, the sender can also remove bias. In this section we investigate that matter. Our results are less general than those in the previous section: we present two different `unbiasing' protocols for parameters $(3,3,1)$. To investigate unbiased announcements, we have over 100 years of design theory to comfortably fall back on. But the investigation of unbiased protocols to produce card deal announcements has not been investigated in a combinatorial setting, as far as we know.

\nl Given parameters $(3,3,1)$, consider again the five-hand announcement $\{012, 034, 056, 135, 246 \}$. There are 60 different five-line announcements containing an arbitrary actual hand \citep[p.56]{ditmarsch:russiancards2003}.\footnote{One of the seven points has to occur thrice in the announcement. In case this is one of the {\bf three} actual cards, one of the three lines containing it will be the actual hand, the four remaining points are distributed over the other two of the three. Given an assignment of any of those four, we can choose one of the remaining {\bf three} to match it. That determines the third of those lines too. Suppose that $i$ is the chosen actual card, $j,k$ the other actual cards, and that the other two lines containing it are $ilm$ and $ino$. Now consider the two lines not containing $i$. One will contain $j$, the other $k$. For the line containing $j$ we can choose one (out of {\bf two}) $l,m$ and one (out of {\bf two}) $n,o$. That determines the fifth hand too. Altogether: $3\cdot3\cdot2\cdot2 = 36$. Else, in case the triple point occurrence is not an actual card, but one of the {\bf four} other points; say $l$. This fixes the lines not containing that point: one of those is now the actual hand, say $ijk$ again, and the other contains the remaining three points, $m, n, o$. Consider the three lines containing $l$. Points (actual cards) $i,j,k$ must be in three different lines containing $l$. For any of those, we can now choose between {\bf 3} of the remaining points $m,n,o$, and for another of those, between {\bf two} of the points still remaining after that choice. Altogether: $4\cdot3\cdot2 = 24$.} In a five-line announcement exactly one point will occur thrice. Suppose 012 is the actual hand. Of the 60 announcements containing 012, 36 contain an actual card 0, 1, or 2 thrice; the remaining 24 therefore contain another card three times. Therefore, a point occurring thrice in this five-line announcement is more likely to be an actual card. A protocol randomly selecting an announcement containing the actual hand therefore propagates this bias, and could rightfully be called a biased protocol. We now adjust (`debias') the protocol as follows (`choose' always means `randomly choose'): 

\begin{fact}[Unbiased five-hand announcement]
Given are parameters $(3,3,1)$. Given an actual hand, choose one among the 36 five-hand announcements containing an actual card thrice, and choose one among the 24 not containing an actual card thrice. Now choose between those two. Given this protocol, there is no relation between the triple point occurrence in the announcement and the actual cards.
\end{fact}

Given that one point occurs thrice in a five line announcement, and given that this is supposed to be meaningless information, sender Alice might as well make {\em public} which point that will {\em always} be, before being dealt a hand of cards, and then execute some protocol resulting in an announcement containing that actual hand, and the pre-announced point thrice, whether it is in the actual hand or not.\footnote{Suggested by Ron van der Meyden} Unfortunately, if we then choose among all such lines, the exact same bias as before again results: 

Given an arbitrary point and an arbitrary line (actual hand), the probability that that point avoids that line is $\frac{6}{7}\cdot\frac{5}{6}\cdot\frac{4}{5} = \frac{4}{7}$, so that the probability that the point is in the line, is $\frac{3}{7}$. There are twelve announcements where the pre-announced point is an actual card, and six where this is not the case.\footnote{In the first case, as before, the four remaining points are distributed over the other two of the three: $3\cdot2\cdot2 = 12$. Else, also as before, the three actual cards {\em must} be in the three different lines containing the preselected point, and for any of those, our options are: $3\cdot2 = 6$.}$^,$\footnote{Let 0 be the publicly known thrice occurring point. In the first case, let 012 be the arbitrary line containing 0. The twelve announcements are: 

\begin{tabular}{l}
012 034 056 135 246 \ \ \ \ 012 034 056 136 245 \\
012 034 056 145 236 \ \ \ \ 012 034 056 146 235 \\
012 035 046 134 256 \ \ \ \ 012 035 046 136 245 \\
012 035 046 145 236 \ \ \ \ 012 035 046 156 234 \\
012 036 045 134 256 \ \ \ \ 012 036 045 135 246 \\
012 036 045 146 235 \ \ \ \ 012 036 045 156 234
\end{tabular}

\noindent On the other hand, let 135 be the arbitrary line not containing 0. Note that this fixes the other line not containing 0. The 6 announcements are:

\begin{tabular}{l}
012 034 056 135 246 \ \ \ \ 012 036 045 135 246 \\
014 023 056 135 246 \ \ \ \ 014 025 036 135 246 \\
016 023 045 135 246 \ \ \ \ 016 025 034 135 246
\end{tabular}
} As $12 \cdot\frac{3}{7} = \frac{36}{7}$, and $6 \cdot\frac{4}{7} = \frac{24}{7}$, our bias is again as before: the odds are 3 to 2 that a point occurring in an announcement is an actual card. But again, we can adjust the protocol, a bit differently now: 

\begin{fact}[Unbiased five-hand announcement with special point]
Given are parameters $(3,3,1)$. Choose one among the twelve possible announcements if the selected thrice occurring point is an actual card, and choose one among the six announcements that are possible if this is not the case. Choose between those two with probability $\frac{4}{7}$ for the first and $\frac{3}{7}$ for the second.
\end{fact}

We close with an additional observation on the status of such protocols. If they are {\em public}, the combination of the protocol and a resulting announcement makes that announcement unbiased {\em for an eavesdropper} with regard to single point occurrence. If they are not public, but, for example, only known between sender and receiver, the situation becomes much more complex. For example, in the absence of information to the contrary, the eavesdropper may incorrectly assume that each line in an announcement equally likely, and from that (correctly) infer that a thrice occurring point is therefore more likely to be an actual card. But this conclusion is then false. Also, if the sender assumes that the eavesdropper follows that line of argument, it would even make sense not to apply an unbiased protocol, but one that is even biased the other way, namely towards triple occurrence of points that are {\em not} actual points. Then again, the eavesdropper may anticipate such behaviour of the sender, etc. In other words, the optimal strategies for sender and eavesdropper under conditions where announcements are always truthful but knowledge of applied protocols is incomplete, are unclear.

On the other hand, incomplete knowledge of a protocol is an unreasonable assumption in our current setting: given the `worst case' assumption where eavesdroppers intercept the entire communication, in other words, where it is a public communication, we might as well assume the `worst case' concerning protocol knowledge: the protocol is public.

\section{Conclusions} \label{furtherwork}

We outlined the need for stricter requirements for cryptographic protocols inspired by the Russian cards problem. A new requirement CA4 has been proposed. This is shown to be equivalent to an alternative requirement CA5. All announcements found to satisfy CA4 are $2$-designs. We have also shown that all binary designs are $3$-designs. Instead of avoiding bias in announcements produced by such protocols, one may as well apply unbiased protocols such that patterns in announcements become meaningless. We gave two examples of such protocols for card deal parameters $(3,3,1)$.

\bibliographystyle{otago}
\bibliography{sigrid,biblio2007}

\end{document}